\documentclass[10pt,prd,superscriptaddress,twocolumn]{revtex4-2}

\usepackage[utf8]{inputenc}
\usepackage{geometry}
\usepackage{mathtools}
\usepackage{amsfonts}
\usepackage{dsfont}
\usepackage{mathrsfs}
\usepackage{bbm}
\usepackage[normalem]{ulem}
\usepackage{slashed}
\usepackage{tensor}

\usepackage{pifont}

\usepackage{adjustbox}

\usepackage{graphicx}
\usepackage{array}[=2016-10-06]
\usepackage{subcaption}
\usepackage{caption}
\usepackage{placeins}
\usepackage{makecell}
\usepackage{float}

\usepackage{xspace}
\usepackage{xfrac}
\usepackage{hyperref}
\usepackage[nameinlink]{cleveref}
\usepackage{appendix}
\usepackage{siunitx}

\Crefname{appsec}{Appendix}{Appendices}

\usepackage[hyphenation]{impnattypo}
\usepackage[all]{nowidow}
\usepackage{microtype}

\usepackage{xifthen}
\usepackage{booktabs}
\usepackage{multirow}
\usepackage[dvipsnames]{xcolor}
\hypersetup{
	colorlinks,
	linkcolor={red!75!black},
	citecolor={blue!75!black},
	urlcolor={blue!75!black},
	pdftitle={Inhomogeneous instabilities in high-density QCD},
	pdfauthor={Pawlowski, Rennecke, Sattler},
}
\usepackage{enumerate}

\usepackage{tikz}
\usetikzlibrary{shapes,arrows,positioning}
\tikzstyle{block} = [rectangle, draw, fill=white!20, 
	text width=10em, text centered, rounded corners, minimum height=2em]
\tikzstyle{dblock} = [draw=black!20!white, text width=10em, dash pattern=on 1pt off 4pt on 6pt off 4pt,
            rectangle, rounded corners, minimum height=2em, text centered]

\newcommand{\imag}{\text{i}}

\newcommand{\LEGO}{LEGO\textsuperscript{\textregistered}}

\graphicspath{{./figures/}}

\newcommand{\gettitle}{Inhomogeneous instabilities in high-density QCD}

\newcommand{\orcid}[1]{\href{https://orcid.org/#1}{\includegraphics[height=1.9ex,width=1.9ex]{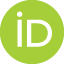}}}

\begin{document}

\title{\gettitle}

\author{Jan~M.~Pawlowski \orcid{0000-0003-0003-7180}\,}
\affiliation{Institut für Theoretische Physik, Universität Heidelberg, Philosophenweg 16, 69120 Heidelberg, Germany}
\affiliation{ExtreMe Matter Institute EMMI, GSI Helmholtzzentrum für Schwerionenforschung mbH, Planckstr.\ 1, 64291 Darmstadt, Germany}

\author{Fabian Rennecke \orcid{0000-0003-1448-677X}\,}
\affiliation{Institut f{\"u}r Theoretische Physik, Justus-Liebig-Universit{\"a}t Gießen, 35392 Gießen, Germany}
\affiliation{Helmholtz Research Academy Hesse for FAIR (HFHF), Campus Gießen, Gießen, Germany}

\author{Franz R. Sattler \orcid{0000-0003-1744-9456}\,}\thanks{\url{fsattler@physik.uni-bielefeld.de}}
\affiliation{Fakult{\"a}t f{\"u}r Physik, Universit{\"a}t Bielefeld, D-33615 Bielefeld, Germany}

\begin{abstract}
	QCD at large densities exhibits a moat regime in the scalar-pseudoscalar sector. The resolution of its dynamics is pivotal for the access to the onset of new phases including the potential critical endpoint of QCD. In this work we present the first selfconsistent analysis of this regime with the functional renormalisation group approach to QCD. We map out the moat regime, including a first analysis of potential inhomogeneous instabilities at baryon chemical potential $\mu_B\gtrsim 600$\,MeV on the  chiral crossover line.   
\end{abstract}

\maketitle

\section{Introduction}
\label{sec:Introduction}

In the past decade, the QCD phase structure at large densities has been mapped out increasingly well with first principles functional QCD, for state-of-the-art results see \cite{Fu:2019hdw, Gao:2020fbl, Gunkel:2021oya}: By now, functional QCD provides quantitative results for the chiral crossover regime with $\mu_B /T \lesssim 4$. For this range of chemical potentials, a critical end point (CEP) could be excluded. 
Moreover, the emergence of such an end point due to chiral fluctuations could be excluded for $\mu_B /T \lesssim 5.5$. In turn, the persistence of a mere chiral crossover can be excluded for $\mu_B/T\gtrsim 7$, which is a very conservative estimate.
As a consequence, the regime $5.5 \lesssim \mu_B /T \lesssim 7$ hosts the onset of new phases (ONP) either in the shape of a CEP or other phenomena.

The systematic theoretical uncertainties in the regime $4 \lesssim \mu_B /T \lesssim 7$ have been narrowed down to the demand of a quantitative resolution of the moat regime in QCD \cite{Pisarski:2021qof, Rennecke:2025bcw}, which has been found in scalar and pseudoscalar correlation functions in \cite{Fu:2019hdw, Fu:2024rto}. In this regime, static mesonic dispersions are minimal at nonzero spatial momentum. This indicates the presence of spatial modulations, with potentially important implications for the phase structure. For example, the conventional homogeneous chirally broken phase, including the CEP, could be unstable against the formation of an inhomogeneous phase at large chemical potential~\cite{Buballa:2014tba}. 
This situation asks for a refined analysis of the crossover regime with $\mu_B /T \gtrsim 4$ in view of the moat behaviour and potential instabilities \cite{Fu:2024rto}. At present, such an analysis can only be done with functional QCD: the moat and even more so the instability are emergent phenomena at finite baryon chemical potential that cannot be extrapolated reliably from QCD at $\mu_B=0$. 
The current work makes an important step in this direction. We resolve the moat regime for $\mu_B/T\gtrsim 4$ for the first time selfconsistently within the functional renormalisation group (fRG) approach to QCD.

\begin{figure*}[ht]
	\centering	
	\includegraphics[width=0.74\linewidth]{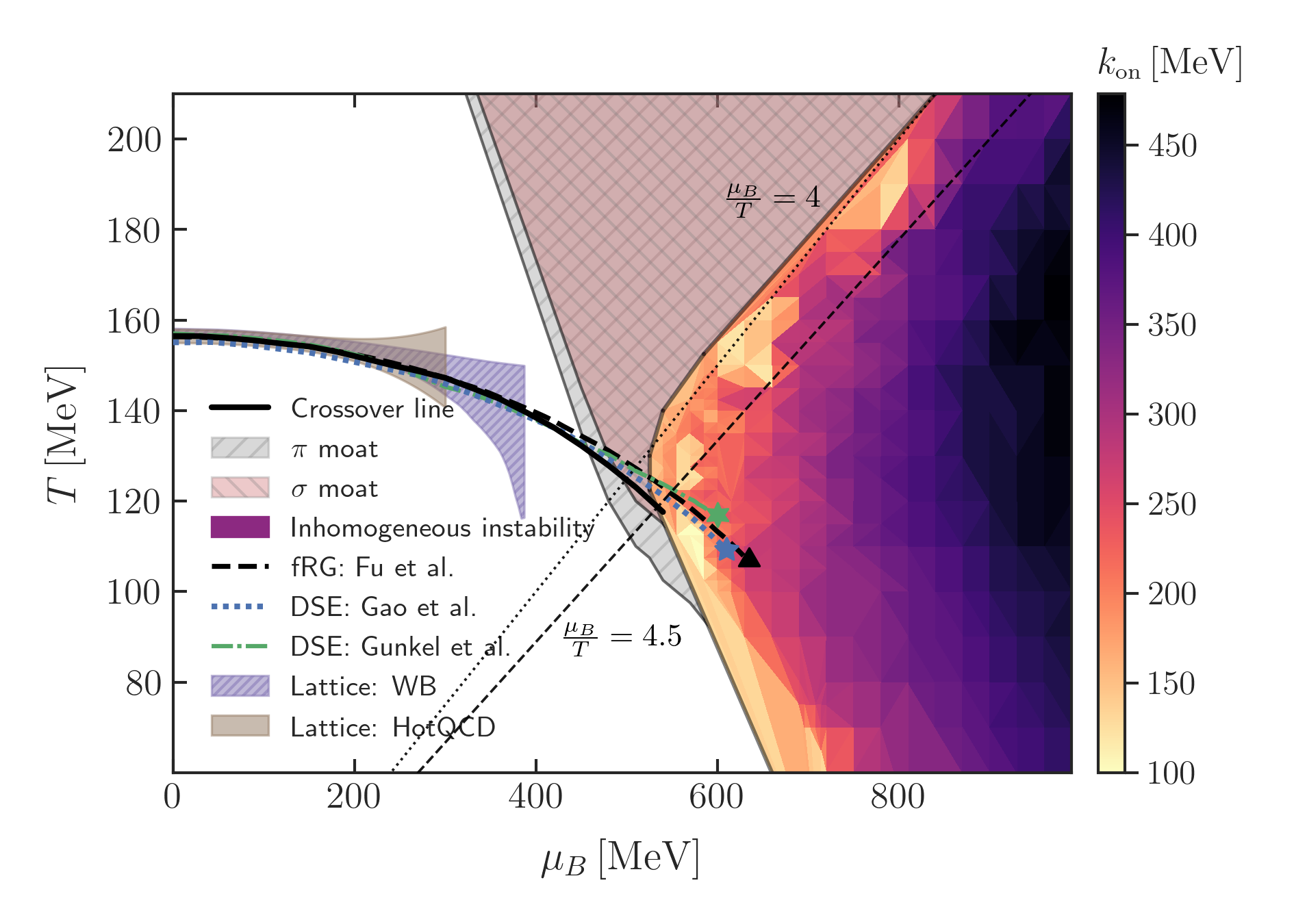} 
    \vspace{-.3cm}
	\caption{
    QCD Phase structure of 2+1 flavour QCD: Our results for the chiral crossover is depicted by the red straight line. We also display the moat regimes of the pions (hatched grey) and the sigma mode (hatched red). The region with signatures of inhomogeneous condensation is shown with a heatmap whose colour indicate value of $k_\textrm{on}$ which is the lowest value of the RG-scale that can be reached before the instability terminates the flow. These instabilities emerge at intermediate length scales. Whether or not this persists up to macroscopic scales, implying inhomogeneous (quasi-) long-range order, requires further studies. The current computation pushes the quantitative reliability bound of functional QCD to $\mu_B/T\approx 4.5$ (dashed black line) with 10\% accuracy. We also show the previous bound $\mu_B/T\approx 4$ (dotted black line). \newline 
	For comparison we also show a compilation of state-of-the-art functional and lattice QCD results: Black dashed line \cite{Fu:2019hdw} (fRG, fQCD), \cite{Gao:2020fbl} (DSE, fQCD), dashed green line \cite{Gunkel:2021oya} (DSE). Violet area \cite{Bellwied:2015rza} (lattice, WB), brown area \cite{Bazavov:2018mes} (lattice, HotQCD).}
	\label{fig:PhaseDiagram}
\end{figure*}

The resulting phase diagram corroborates the result in \cite{Fu:2019hdw} and is displayed in \Cref{fig:PhaseDiagram}. We also show that potential instabilities emerges for $\mu_B/T\gtrsim 5.5$ at intermediate momentum scales. This may also indicate a precondensation phenomenon \cite{Khan:2015puu, PGPS2026} in this regime.
Importantly, the instability regime covers the potential CEP, as estimated with functional QCD for $5.5\lesssim \mu_B/T\lesssim 6.2$, \cite{Fu:2019hdw, Gao:2020fbl, Gunkel:2021oya}. A comprehensive analysis of this regime, including the persistence of the instability for large volumes, will be presented elsewhere.

%

\section{Functional QCD}
\label{sec:fRG}

Assessing the physics and dynamics in the QCD moat regime requires the full spatial momentum resolution of the scalar-pseudoscalar four-quark scattering channel.
In the fRG approach with emergent composites \cite{Gies:2002hq, Braun:2009ewx, Mitter:2014wpa, Braun:2014ata, Rennecke:2015eba, Fu:2019hdw, Ihssen:2024miv, Fu:2024rto}, as used for phase structure investigations this is tantamount to resolving the full spatial momentum dependence of the mesonic pion and $\sigma$-mode two-point functions.
In the aforementioned works, the momentum dependence of the mesonic propagators has been approximated with an RG-scale dependent wave function. This has been proven to be quantitatively accurate (on the percent level) for monotonic dispersion of pions in \cite{Helmboldt:2014iya}, which also informed our systematic error estimates.

As has been observed for the first time in \cite{Fu:2019hdw}, close to the chiral crossover line, the QCD moat emerges for $\mu_B/T\gtrsim 4$, see also \cite{Fu:2024rto}.
We note in passing that the remaining source of the increasing systematic error in functional QCD computations for 
$\mu_B/T\gtrsim 4$ has been investigated in a companion work \cite{WZHWYF2026} in the fQCD collaboration \cite{fQCD}: for $\mu_B/T\lesssim 4$ the additional channels do not play a rôle at $\mu_B/T\lesssim 7$ and can be safely dropped. This confirms the qualitative picture already discussed in \cite{Braun:2019aow}. For $\mu_B/T\lesssim 7$ the diquark channel becomes relevant, see \cite{Braun:2019aow}. 

In \cite{Fu:2024rto} the non-trivial spatial momentum dependence was further analysed by \textit{reading out} the full complex frequency and spatial momentum dependence. This provides access to the spectral properties in the QCD moat regime. However, this frequency and momentum dependence was not fed back into the diagrams. With the deepening of the moat, and even more so in a potentially inhomogeneous regime, such a feed back is key to the reliability of the results. 

In the present work we implement the full spatial momentum dependence of the emergent pions and $\sigma$-mode selfconsistently: it is fed back into the flow diagrams which allows us to study the change of the dynamics caused by the successive deepening of the QCD moat.  
Our current truncation builds on previous functional QCD calculations for the phase diagram \cite{Fu:2019hdw, Ihssen:2024miv}, with several key improvements. 
\begin{itemize}
	\item We also introduce emergent composites (dynamical hadronisation) for a part of the strange-quark sector. Specifically, we use an emergent composite for the scalar channel of the four-$s$ scattering vertex. This channel contributes to the strange mass function and encodes the momentum-dependent change from the current to the constituent strange mass. Although this has a minimal impact on the QCD dynamics in comparison to \cite{Fu:2019hdw, Ihssen:2024miv, WZHWYF2026}, it facilitates the computation of the full momentum-dependent strange quark propagator.
	\item To improve the systematics in the region of large $\mu_B$, we resolve the full field-dependence of $V(\phi_l^2)$. This implies that we resolve multi-meson scatterings to arbitrary order. Explicitly, we use finite element techniques, as previously developed for the fRG in \cite{Grossi:2019urj, Grossi:2021ksl, Ihssen:2022xkr, Ihssen:2023qaq, Ihssen:2023xlp, Ihssen:2024miv}
	\item We resolve the dependence of the meson propagators $G_{\phi\phi}$ (with $\phi=\pi,\,\sigma$) and quark propagators $G_{\bar{q}q}$ on the spatial momentum $\boldsymbol{p}$ at all RG scales $k$, and feed them back into the flow. This allows us for the first time to resolve the moat regime in a fully selfconsistent manner.
	\item The glue sector is calculated with an identification of the RG-scale and the physical momentum scale~${k=p}$. We calculate temperature- and quark-dependent corrections for glue and ghost propagators around vacuum data taken from \cite{Cyrol:2017ewj}. Other gluonic correlation functions are computed selfconsistently for all  temperatures. The quantitative nature of this approximation has been checked thoroughly in \cite{Fu:2019hdw, Ihssen:2024miv} and is supported by the \LEGO-principle \cite{Ihssen:2024miv}, see also the discussion below. 
    The gluon propagator is parametrised by using a flowing gluon mass gap, which carries most of the thermal effects of the gluon propagator. The resulting propagator fits the lattice data from \cite{Ilgenfritz:2017kkp} very well, see \Cref{fig:gluonProps}. 
	\item The Polyakov-loop potential $V_\textrm{glue}(A_0)$ is evaluated similarly to \cite{Fu:2019hdw}: we take pure Yang-Mills input from \cite{Lo:2013hla} and add to it the quark-corrections from our flows.
\end{itemize}
For more details on the computation see \Cref{app:setup}, as well as \cite{Sattler:2025hcg}. We finalise this discussion of the setup with one of the systematic error estimate: This analysis is based on the conceptual approach to systematic error control in functional approaches detailed in \cite{Ihssen:2024miv}: We have calculated our results with three different regulators, see \Cref{fig:QCDPD_Regulators}. These regulators are rather different and lead to a comprehensive check of the regulator dependence or rather the lack thereof. Additionally, we have performed a calculation with a gluon propagator which is modified within the bounds of the error estimate in \cite{Cyrol:2017qkl}, see also \cite{Ferreira:2025tzo} for a systematic analysis of the respective error estimates. This exploits the modular form of flow equations (\LEGO principle) to infer error propagation from the high-energy sector to the low-energy observables. This leads us to improved quantitative reliability bound of functional QCD in the vicinity of the chiral crossover line of
\begin{align}
    \mu_B/T \approx 4.5\,,
\end{align}
see \Cref{sec:Instability} and \Cref{fig:PhaseDiagram}.

\begin{figure}[t]
	\centering	
	\includegraphics[width=0.95\columnwidth]{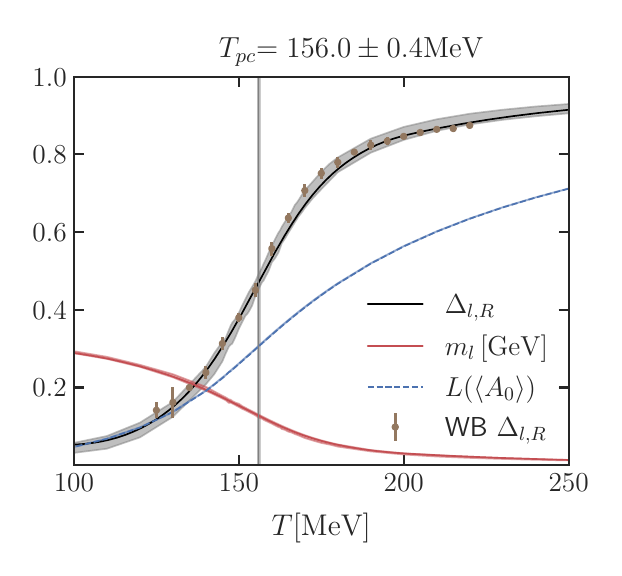}
	\caption[]{The reduced light condensate $\Delta_{l,R}$  \labelcref{eq:Delta}, light quark mass $m_l$ and Polyakov loop expectation value $L(A_0)$ along the temperature axis. The pseudo-critical temperature $T_{pc}$ is indicated together with its error band. We compare this to lattice data from the Wuppertal-Budapest collaboration \cite{Borsanyi:2010bp}.}
	\label{fig:QCDPD_T_axis}
\end{figure}

We close this Section with a brief discussion of the benchmark results: we have computed the full flows for discrete points within the range $\mu_B \in [0,1000\,\textrm{MeV}]$ and $T\in[0,300\,\textrm{MeV}]$. The resulting phase diagram is shown in \Cref{fig:PhaseDiagram}, and key quantities on the $\mu_B=0$ axis are shown in \Cref{fig:QCDPD_T_axis}.
From our calculation, we obtain a pseudo-critical temperature at $\mu_B = 0$.
\begin{align}
	T_{pc} = 156.0 \pm 0.4\,\textrm{MeV}\,,
\end{align}
and a curvature of
\begin{align}
	\kappa = 0.0152(22)\,.
\end{align}
As seen in \Cref{fig:PhaseDiagram} and \ref{fig:QCDPD_T_axis}, our results are in quantitative agreement with the available lattice QCD results at $\mu_B=$ and their finite density extrapolations, see e.g.~\cite{Bellwied:2015rza, Bazavov:2018mes} as well as state-of-the-art functional QCD results \cite{Fu:2019hdw, Gao:2020fbl, Gunkel:2021oya}. A respective compilation of benchmark results can be found in \cite{FunReview2025}.

\section{Moat regime}
\label{sec:Moat}

In the moat regime, the static two-point function of a boson $\phi$, e.g.~a meson in QCD, has a global minimum at nonzero spatial momentum $\boldsymbol{p}_{M_\phi}>0$ \cite{Pisarski:2021qof},
\begin{align}
   \min_{\boldsymbol{p}} \Gamma_{\phi\phi}(\boldsymbol{p}) = \Gamma_{\phi\phi}(\boldsymbol{p}_{M_\phi})\,.
\label{eq:motedef}
\end{align}
Hence, the screening potential between quark/nucleon pairs features spatial modulations with wave number $\boldsymbol{p}_{M_\phi}$ if it receives contributions from the static meson propagator $G_{\phi\phi} = \Gamma_{\phi\phi}^{-1}$. Specifically, this happens for $\phi=(\sigma,\boldsymbol{\pi})$. 

\Cref{eq:motedef} also implies that the spectral function of $\phi$ is enhanced in the spacelike region \cite{Fu:2024rto}. This is a consequence of Landau damping: the spectral weight in the spacelike region stems from the decay/absorption of a virtual $\phi$ into real particle-hole pairs in the medium, and the phase space for this process is enhanced around $\boldsymbol{p}_{M_\phi}$ in the moat regime. The corresponding collective spacelike mode is called a moaton \cite{Fu:2024rto}.

The moat regime for $\phi$ is split into two distinct sub-regimes, characterised by $\Gamma_{\phi\phi}(\boldsymbol{p}_{M_\phi}) > 0$ and $\Gamma_{\phi\phi}(\boldsymbol{p}_{M_\phi}) \leq 0$:\\[-1ex]

For $\Gamma_{\phi\phi}(\boldsymbol{p}_{M_\phi}) > 0$, the system is in a stable homogeneous phase. Many available results indicate that this phase is disordered, with propagators decaying like a modulated exponential \cite{Fu:2019hdw, Koenigstein:2021llr, Pannullo:2024sov, Topfel:2024iop, Cao:2025zvh, Motta:2025xop, Rennecke:2025kub}, see also \cite{Diaz-Alonso:1998eva, Liu:2007bu, Chakrabarty_2011, Schindler:2019ugo, Pisarski:2020dnx, Schindler:2021otf} for related results. This may be understood on the example of QCD in the chiral limit, where pions are massless in the ordered hadronic phase, $\Gamma_{\pi\pi}(0)=0$. \Cref{eq:motedef} would then imply a negative two-point function at $\boldsymbol{p}_{M_\phi}^2 > 0$, contradicting the initial assumption. The same argument applies to the critical mode of the CEP \cite{Haensch:2023sig}.\\[-1ex]

For $\Gamma_{\phi\phi}(\boldsymbol{p}_{M_\phi}) \leq 0$, the system is unstable \cite{Weinberg:1987vp}. At $\Gamma_{\phi\phi}(\boldsymbol{p}_{M_\phi}) = 0$, the moaton develops into a massless static mode with nonzero momentum. The system hence becomes unstable and wants to condense the moatons into an ordered, inhomogeneous phase with nonzero wavenumber $\boldsymbol{p}_{M_\phi}$ \cite{Fu:2024rto}. We refer to \cite{Motta:2023pks, Motta:2024agi, Motta:2024rvk, Fu:2024rto} for techniques to detect such instabilities in QCD.\\[-1ex] 

\begin{figure*}[t]
    \includegraphics[width=0.495\linewidth]{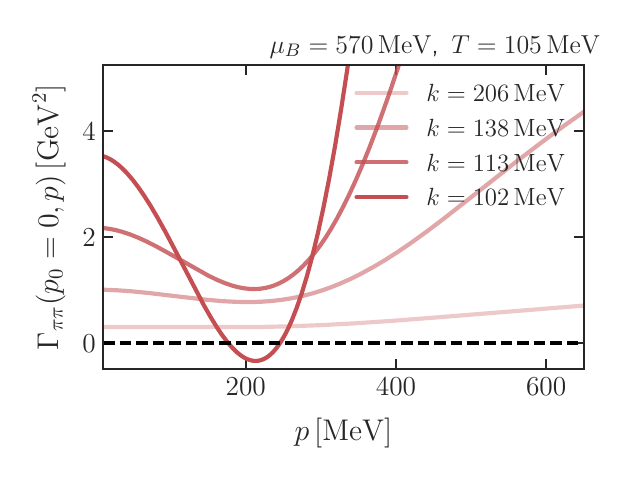}
    \includegraphics[width=0.495\linewidth]{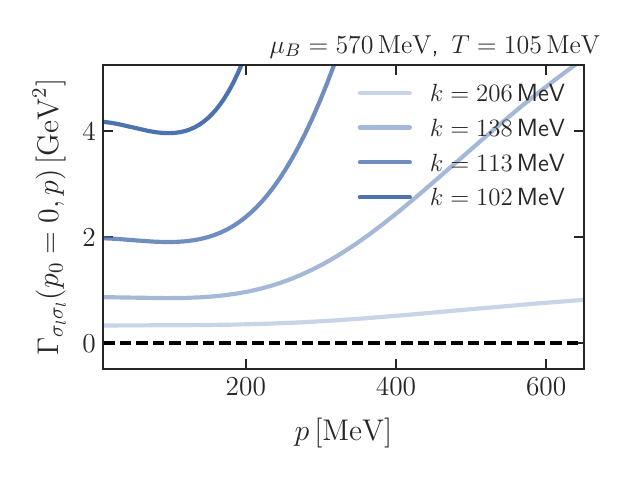}

    \includegraphics[width=0.495\linewidth]{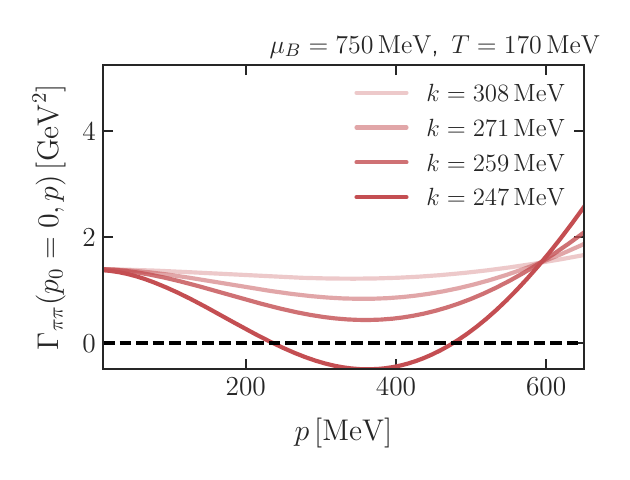}
    \includegraphics[width=0.495\linewidth]{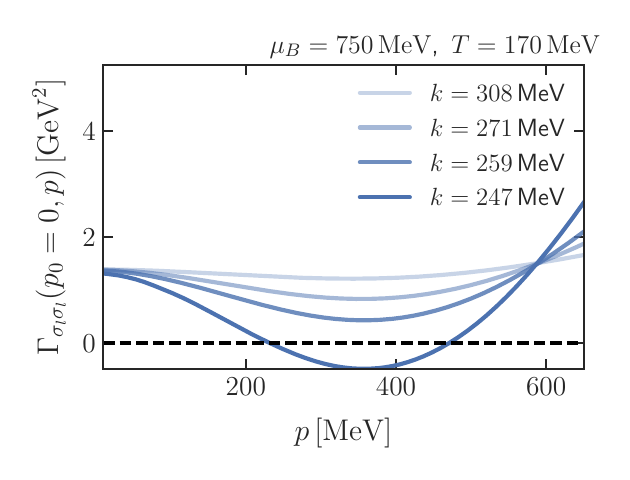}
    \caption{Static pion and sigma two-point functions at different cutoff scales. The upper row shows the instability in the vicinity of the chiral crossover line. The lower row is at higher temperatures deep in the chirally restored~phase.}
    \label{fig:Ginst}
\end{figure*}
This classification in terms of the sign of $\Gamma_{\phi\phi}(\boldsymbol{p}_{M_\phi})$ becomes intricate if the system features algebraic (quasi-long-range) order. In this case, there is a massless mode giving rise to an infinite correlation length, but, owing to strong fluctuations of this mode, the order parameter correlation never develops an instability, i.e.~$\Gamma_{\sigma\sigma}(\boldsymbol{p}_{\sigma}) > 0$ if $\sigma$ is the order parameter field. This can arise, for example, from a Landau-Peierls instability of a would-be inhomogeneous phase, see e.g.~\cite{Lee:2015bva, Hidaka:2015xza} for studies in the QCD context. Then, the resulting phase is some type of liquid crystal, and the situation is reminiscent of the Berezinskii–Kosterlitz–Thouless (BKT) phase of systems with continuous symmetry in two spatial dimensions \cite{Berezinskii:1970pzv, Kosterlitz:1973xp}. This highlights an important feature relevant for the interpretation of our results: phases with algebraic order look ordered at intermediate scales, but not in the deep infrared (IR). 
While not fully settled yet, this might be encoded in the precondensation phenomenon in the fRG framework, where a nonzero order parameter develops at intermediate RG scales $k$, but vanishes again in the IR for $k\rightarrow 0$; see, e.g., \cite{Khan:2015puu, PGPS2026} for the QCD context and \cite{VonGersdorff:2000kp, Jakubczyk:2014isa, Jakubczyk:2016rvr, Tolosa-Simeon:2025fot} for BKT phases.
In practice, we expect an instability $\Gamma_{\phi\phi, k>0}(\boldsymbol{p}_{M_\phi}) = 0$ in some finite $k$ interval, while $\Gamma_{\phi\phi, k=0}(\boldsymbol{p}_{M_\phi}) > 0$.

In summary, the moat behaviour describes various spatially modulated regimes, including modulated disordered and inhomogeneous phases, but also algebraic order. Which of these regimes is realized, is ultimately decided in the deep IR. Importantly, an instability at intermediate scale $k = k_{\rm on} > 0$ suggests the onset of either inhomogeneous (long-range) or algebraic (quasi-long-range) order.  

\begin{figure*}[t]
	\centering
    \begin{minipage}{0.48\linewidth}
    	\includegraphics[width=1\columnwidth]{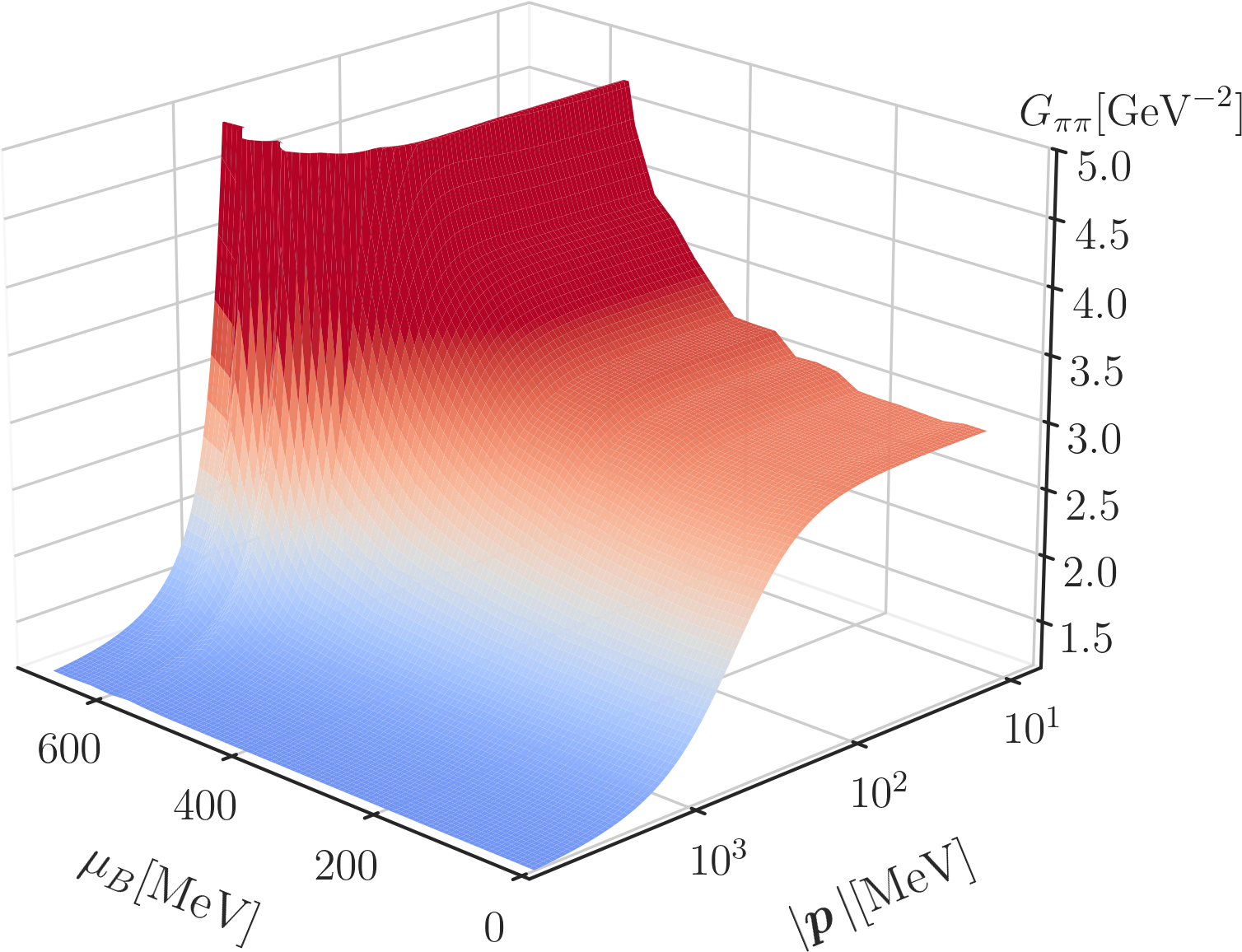}
    	\caption[]{The pion propagator $G_{\pi\pi}(\boldsymbol{p})$ along the crossover line and inside the instability region. The values of $T$ associated with the respective $\mu_B$ are shown in the inlay plot. Note that the onset of the instability leads to a divergence in $G_{\pi\pi}(|\boldsymbol{p}|\approx 200\,\textrm{MeV})$, which at this point is no longer the propagator as found in the ground state of the theory. The shown propagator is normalised to 1 at $|\boldsymbol{p}| = 5\,\textrm{GeV}$.}
	\label{fig:pionProp}
    \end{minipage}
    \hspace{0.02\linewidth}
    \begin{minipage}{0.48\linewidth}
    	\includegraphics[width=1.04\columnwidth]{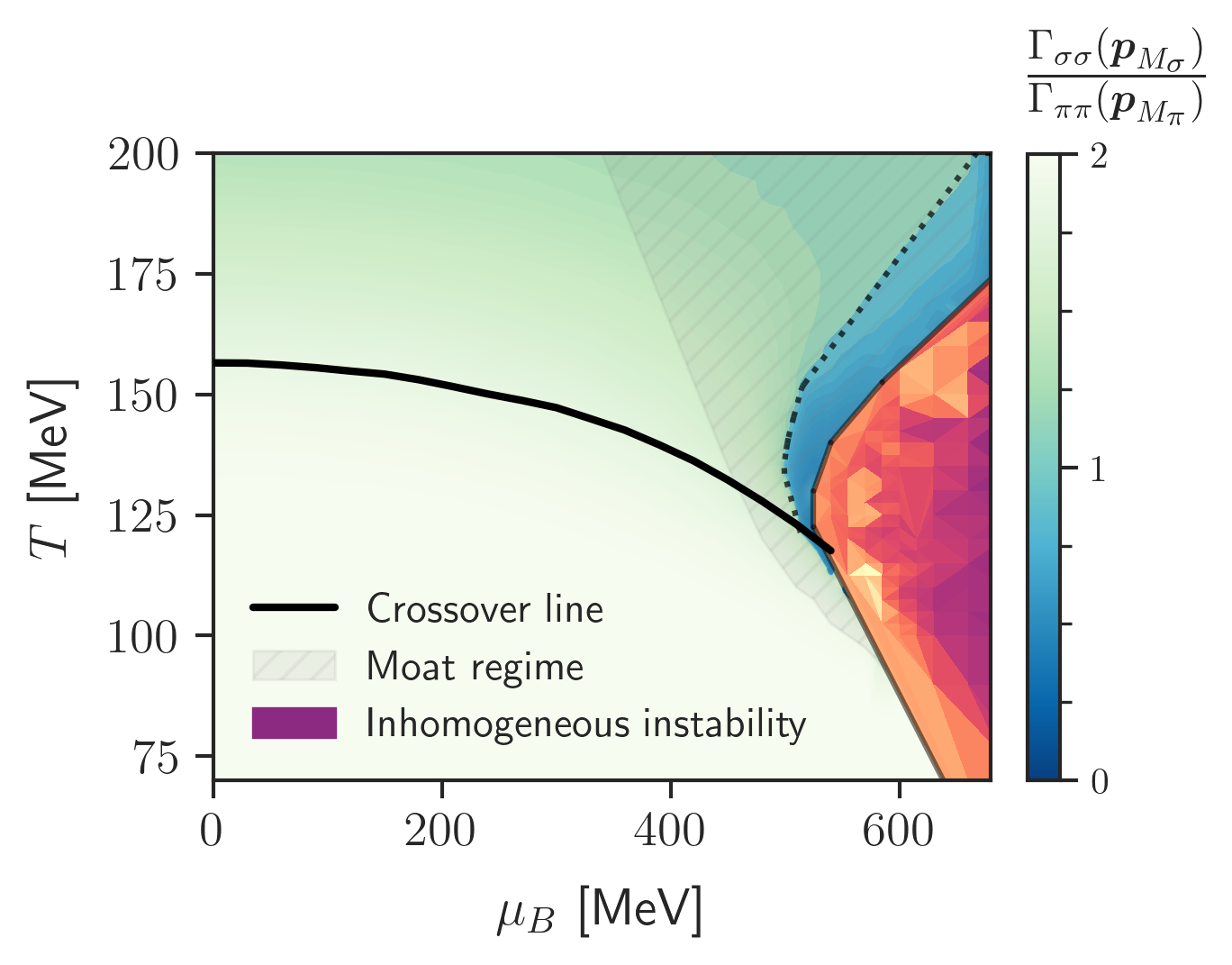}
    \caption{Heatmap of the ratio $\Gamma_{\sigma\sigma}(\boldsymbol{p}_{M_\sigma})/\Gamma_{\pi\pi}(\boldsymbol{p}_{M_\pi})$, evaluated at their respective  minima $\boldsymbol{p}_{M_\sigma},\,\boldsymbol{p}_{M_\pi}$, see \labelcref{eq:motedef}. This ratio is a rough estimate for that of the correlation lengths $\xi_\sigma/\xi_\pi$. The dotted line indicates where the ratio becomes 1.}
    	\label{fig:Softestmode}
        \vspace{11.9ex}
    \end{minipage}
\end{figure*}
%

\subsection{Emergent instabilities}
\label{sec:Instability}

As shown in \Cref{fig:PhaseDiagram}, the moat regime persists also in light of the systematic improvements made in this work. This further corroborates its fundamental importance for the QCD phase diagram and the location of the ONP.
In addition, we find, for the first time in QCD, the instabilities discussed before.

As an example, we show the static pion and sigma two-point functions at $T = 105$\,MeV and $\mu_B=570$\,MeV in the upper row of \Cref{fig:Ginst}. The static two-point functions start to develop a moat behaviour with a nonzero energy gap at $k\approx 200$\,MeV. Towards smaller $k$, the moat deepens and the gap gets progressively smaller. At the onset scale $k_{\rm on}\approx 100$\,MeV, the pion develops an instability at $\boldsymbol{p}_{M_\pi} \approx 200$\,MeV. 
The same happens also deep in the chirally restored phase at $T = 170$\,MeV and $\mu_B=750$\,MeV, see the lower row of \Cref{fig:Ginst}. In this case, the moat and the instability develop at larger scales. 
At $k_{\rm on}$, we terminate the RG flow, because the instability suggests the formation of an intermediate inhomogeneous phase and hence the expansion of the RG flow around a respective background. Here, the deep IR remains inaccessible as we focus on homogeneous phases and their instabilities. We hence cannot answer whether there is modulated long-range order, quasi-long range order, or disorder at this point. But our results at least suggest the presence of either an inhomogeneous or a liquid-crystal like phase in QCD.

We find a large region with such instabilities: the homogeneous moat regime appears to always become unstable at sufficiently high density.
The heatmap in \Cref{fig:PhaseDiagram} shows the onset scale, $k_{\rm on}$, where the instabilities occur. The larger $\mu_B$, the earlier in the RG flow (i.e.~at larger $k$) the system becomes unstable. Furthermore, the spatial momentum $\boldsymbol{p}_M$, at which the instability occurs, also increases with $\mu_B$ for a given $T$. Hence, the wavenumber of the underlying spatial modulation increases with the density. This is perhaps not surprising, because particle-hole fluctuations around the (thermally smeared) Fermi surface give rise to the moat regime in the first place.    

Following the chiral crossover (the red line in \Cref{fig:PhaseDiagram}), the instability occurs occurs at $\mu_B \approx 570$\,MeV. The static pion propagator along the crossover line is shown in \Cref{fig:pionProp}. The instability manifests in a divergence at nonzero spatial momentum. Importantly, this happens \emph{before} the crossover turns into the CEP. This opens up the possibility that there is no ordinary CEP of the chiral phase transition in QCD. Instead, it may feature a Lifshitz point \cite{Hornreich:1975zz, Nickel:2009ke}, a Lifshitz regime \cite{Pisarski:2018bct} or something entirely unknown.

Interestingly, this intermediate instability in the vicinity of the crossover occurs first for pions, see the upper row of \Cref{fig:Ginst}. However, for sufficiently large chemical potentials $\mu_B\gtrsim 550$\,MeV and close to the crossover line, the $\sigma$-mode becomes lighter than the pion mode. This is illustrated in \Cref{fig:Softestmode}, where we show a heatmap of the ratio $\Gamma_{\sigma\sigma}(\boldsymbol{p}_{M_\sigma})/\Gamma_{\pi\pi}(\boldsymbol{p}_{M_\pi})$, evaluated at their respective  minima, evaluated at their respective  minima $\boldsymbol{p}_{M_\sigma},\,\boldsymbol{p}_{M_\pi}$, see \labelcref{eq:motedef}. This ratio is a rough estimate for that of the correlation lengths, $\xi_\pi/\xi_\sigma$. On the crossover line it drops below 1 for $\mu_B\approx 550$\,MeV, indicating a lighter $\sigma$-mode. Moreover, the ratio gets successively smaller until the boundary of the instability regime is hit at $\mu_B\approx 570 $\,MeV. 
A successively lighter $\sigma$ may indicate the approach towards a regime with the critical $Z_2$-scaling of the CEP. Importantly, in the regime with the intermediate instability, this is undone instantly by the drastic change of the static pion propagator close to the chiral crossover line, see  \Cref{fig:Ginst}. It is very suggestive that the approach to criticality drowns completely in the dynamics of the instability. However,  
we also stress that the critical mode of the CEP is a mixture of the $\sigma$-mode, the density mode and the Polyakov loops \cite{Haensch:2023sig}.
Thus, a direct statement about how the instability affects the CEP cannot be made.

Finally, the onset chemical potential $\mu^\textrm{on}_B$ of the instability region on the crossover line shows a mild regulator dependence: $\mu^\textrm{on}_B \approx 540 - 620$\,MeV, consistent with our systematic error estimate  error of $\sim 10\%$ for the present truncation in this regime. The accuracy of this bound on the ONP region will be improved in a forthcoming work. There we consider further interaction channels, and specifically the diquark and density channels in the four-quark scattering vertex, as well as a better momentum resolution of all correlation functions. 

In summary, with the present analysis we cannot unravel how the critical mode of the CEP is affected by the instability. Note that if the critical mode itself shown moat behaviour, it automatically implies an instability \cite{Haensch:2023sig}. This leaves us with the crucial question about the nature of the ONP: the emergent instability may 'swallow' the CEP or it may persist.  

In any case, our results provide first indications that the ground state of dense QCD matter might not be homogeneous. Its exact nature, and the fate of the CEP remain open questions.
Reliable answers require a quantitative functional QCD analysis, whose setup and foundation is provided in the present work.

\section{Conclusion}
\label{sec:conc}

In the present work we have initiated a comprehensive selfconsistent analysis of the QCD moat regime at large density within functional QCD. In particular, the full frequency- and momentum dependences of the pion and $\sigma$-mode two-point functions have been computed and fed back to the flow diagrams. This led us to the phase structure depicted in \Cref{fig:PhaseDiagram}. 

The results further solidify the current state-of-the art results in functional QCD, \cite{Fu:2019hdw, Gao:2020fbl, Gunkel:2021oya}, and in particular the location of the chiral phase boundary. At chemical potentials $\mu_B/T \gtrsim 4$, QCD exhibits a moat regime with an increasingly deep moat, and for $\mu_B/T \gtrsim 6$, QCD develops an instability at successively larger intermediate momentum scales. This indicates that inhomogeneous or liquid crystal-like phases may form in dense QCD matter. 

It remains to be seen whether this instability survives the vanishing momentum scale limit. A reliable numerical analysis of this chiefly important question requires a further improvement of the current approximation and results will be presented in a forthcoming work. Finally, the present work will allow us to investigate quantitatively experimental signatures of the QCD moat as well as the instability regime, following \cite{Pisarski:2021qof, Rennecke:2023xhc, Nussinov:2024erh}. We also hope to report on respective results in the near future.

\begin{acknowledgments}
We thank Wei-jie Fu, Chuang Huang, Flavio Nogueira, Zohar Nussinov, Michael Ogilvie, Robert Pisarski, Stella Schindler and Shi Yin for discussions and collaboration on related topics. This work is done within the fQCD collaboration \cite{fQCD}.

FRS acknowledges funding by the GSI Helmholtzzentrum f\"ur Schwerionenforschung and by HGS-HIRe for FAIR. FRS is supported by the Deutsche Forschungsgemeinschaft (DFG, German Research Foundation) through the Emmy Noether Programme Project No. 54526179. 
This work is also funded by the Deutsche Forschungsgemeinschaft (DFG, German Research Foundation) under Germany’s Excellence Strategy EXC 2181/1 - 390900948 (the Heidelberg STRUCTURES Excellence Cluster), the Collaborative Research Centre SFB 1225 - 273811115 (ISOQUANT), and the CRC-TR 211 “Strong-interaction matter under extreme conditions” – project number 315477589 – TRR 211.
\end{acknowledgments}

\appendix
  \crefalias{section}{appsec}


\section{Truncation}
\label{app:setup}

In this appendix, we briefly explain certain key points of our setup. As we build on the apparatus of \cite{Ihssen:2024miv}, we refer for details to this work, only pointing out the improvements over \cite{Ihssen:2024miv} here. For further details, see also \cite{Sattler:2025hcg}.

\subsection{Light meson sector}
\label{sec:QCDPD_setup_meson}

We parametrise the pion inverse propagator as 
\begin{equation}
	\Gamma_{\pi\pi}(p_0, \boldsymbol{p}) = Z_\phi^\parallel\,\left[ p_0^2 + z_\phi^\perp(p) \boldsymbol{p}^2 + m_\pi^2 \right]
	\,,
\end{equation}
where we include an additional full momentum dependence in the spatial momentum, allowing to resolve the moat regime selfconsistently.
Furthermore, we simplify our flows for the sigma meson by using the wave functions $Z_\phi^\parallel$ and $z_\phi^\perp(p)$ also there, and only change the sigma meson mass when we couple it back into the dynamics, i.e. we use the Ansatz
\begin{equation}
	\Gamma_{\sigma_l\sigma_l}(p_0, \boldsymbol{p}) = Z_\phi^\parallel\,\left[ p_0^2 + z_\phi^\perp(p) \boldsymbol{p}^2 + m_{\sigma_l}^2 \right]
	\,,
\end{equation}
The wave-function parallel to the heat bath is projected at $p = 0$ from the flow of the pion two-point function, and the corresponding projection operator reads
\begin{equation}
	Z_{\phi}^\parallel = \partial_{p_0^2} \Gamma_{\pi\pi}\Big\vert_{p=0}
	\,.
\end{equation}
We absorb $Z_\phi^\parallel$ using the RG-invariant formulation of the fRG into the mesonic fields, see \cite{Ihssen:2024miv}. $z_\phi^\perp(p)$ is computed in a fully momentum-dependent manner with the spatial projection
\begin{equation}
	z_\phi^\perp(p) = \frac{\Gamma_{\pi\pi}(p) - \Gamma_{\pi\pi}(0)}{\boldsymbol{p}^2}   \Bigg\vert_{p_0=0}
	\,.
\end{equation}
We assume use that $z_{\phi,\perp}(p_0,|\boldsymbol{p}|) \approx z_\phi^\perp(0, \sqrt{p_0^2 + \boldsymbol{p}^2})$, which has been previously used e.g. in \cite{Cyrol:2017qkl}.

To correctly evaluate observables related to the sigma meson, we also read out the actual propagator of the sigma meson, though we do not feed it back into the system. We parametrise it as
\begin{equation}
	\tilde\Gamma_{\sigma_l\sigma_l}(p_0, \boldsymbol{p}) = Z_\phi^\parallel\,\left[ Z_{\sigma_l}^\parallel p_0^2 + z_{\sigma_l}^\perp(p) \boldsymbol{p}^2 + m_{\sigma_l}^2 \right]
	\,.
\end{equation}
The projection follows as for the pion propagator.

\subsection{Quark sector}
\label{sec:QCDPD_setup_quark}

We parameterise the quark two-point function in a fully Lorentz symmetric manner as
\begin{align}
	\Gamma_{q\bar{q}}(p_0, \boldsymbol{p}) = Z_q \big[&
	z_{q}(p) (\gamma_0(p_0 - \imag \mu_q) + \gamma_i p_i) + M_q(p)
	\big]	 \notag\\[1ex]
	&- \imag g_{A\bar{q}q} A_0
	\,,\quad\text{with}\,q=l,s
	\,.
\end{align}
Once again, the wave-function is calculated from its spatial part and we use the approximation ${z_{q}(p_0,|\boldsymbol{p}|) \approx z_{q}(0, \sqrt{p_0^2 + \boldsymbol{p}^2})}$.
The corresponding projection is simply proportional to $\imag\slashed{\boldsymbol{p}}$.
The overall scalar wave-function $Z_q$, projected in the same manner, is absorbed into the quark fields using the {RG}-invariant scheme of \cite{Ihssen:2024miv}.
As a consequence of this parametrisation, $z_{q}(0) = 1$.

The quark mass is composed of the {VEV} of the sigma mode $\sigma_q$ and the Yukawa coupling $h_q(p)$ with $q=l,s$. We project the momentum dependence of the Yukawa couplings again from the spatial part of the quark two-point function, i.e. we take ${h_q(\pi T + p_0,|\boldsymbol{p}|) \approx~h_q(\tilde\nu_0, \sqrt{p_0^2 + \boldsymbol{p}^2})}$. 
Furthermore, we utilise for the frequency part the Silver-Blaze symmetric point and the lowest fermionic Matsubara mode:
\begin{equation}
	\tilde \nu_0 = \alpha(k,T) \pi T - \imag \mu_q
	\,.
\end{equation}
In \cite{Fu:2015naa, Fu:2016tey, Fu:2019hdw}, it has been observed that to suppress the artificial $T$-dependence at high cutoff scales $k$, an exponential suppression $\alpha(k,T)$ of the lowest fermionic Matsubara mode is necessary in this approximation. We take, similarly as in these works,
\begin{equation}
	\alpha(k,T) = e^{-\frac{k}{2\pi T}}
	\,.
\end{equation}
The projection itself is then identical to the one used in the vacuum, for more details see \cite{Ihssen:2024miv}.

Note that for \Cref{fig:QCDPD_T_axis} and \Cref{fig:PhaseDiagram}, we use the reduced light condensate,
\begin{equation}
	\Delta_{l,R}(T,\mu_B) = - \frac{\Delta_{l}(T,\mu_B) - \Delta_{l}(0,0)}{\Delta_l(0,0)}
	\,.
\end{equation}
where $\Delta_q$ is given by
\begin{align}
	\Delta_{q} &= m_q^0\,\frac{\partial \Gamma[\Phi_\textrm{EoM};T,\mu_B]}{\partial m^0_q}
	= m_q^0\,\frac{T}{\mathcal{V}}\int_x \langle \bar q(x) q(x) \rangle
	\notag\\[1ex]
	&=-m_q^0\,\int_x \text{tr}\,G^{q\bar{q}}
	\,,\qquad\text{with}\quad q=l,s.
    \label{eq:Delta}
\end{align}
%

\subsection{Strange meson sector}
\label{sec:QCDPD_setup_strange}

In this work, we dynamically bosonise the scalar strange composite.
The relevant four-fermion interaction for the strange quark mass is
\begin{equation}\label{eq:QCDPD_scalarStrange4F}
	\mathcal{L}_{\sigma_s} = \lambda_{\sigma_s} (\bar{s}s)^2
	\,.
\end{equation}
We refrain from using a full $N_f=2+1$ basis for the projections, which can be found e.g. in \cite{Braun:2020mhk}. Rather, we use the same four-fermion sub-basis for $N_f=2$ as in \cite{Ihssen:2024miv} and add the interaction \labelcref{eq:QCDPD_scalarStrange4F} directly as a further tensor (orthogonal to all others), and require that the rest of the (implicitly defined) $N_f=2+1$ basis is orthogonal to all the elements we just mentioned.
This implies that the projector onto $\lambda_{\sigma_s}$ is directly given by 
\begin{equation}
	\frac{1}{\mathcal{N}_{\sigma_s}}\mathbbm{1}_s\times \mathbbm{1}_s
	\,,
\end{equation}
with $\mathcal{N}_{\sigma_s} = 4N_c$ being the normalisation of the projector.
We proceed by the dynamical hadronisation program as can be also found in  \cite{Fu:2019hdw, Ihssen:2024miv}. The corresponding hadronisation function reads
\begin{align}
	\dot{\sigma_s} &= \int_p\dot{A}_s(p)\,(\bar{s}s)(p)
	\,,
\end{align}
with
\begin{align}
	\dot{A}_s(p) = -\frac{1}{h_s(p)}\text{Flow}(\lambda_{\sigma_s})
	\,,
\end{align}
where we have introduced the scalar strange composite $\sigma_s$ and $h_s(p)$ as the strange Yukawa coupling between the strange composite and the strange quarks.
Correspondingly, the hadronisation function feeds into the flow of $h_s(p)$ with
\begin{align}
	\partial_t h_s(p) = \left(\frac{\eta_{\sigma_s}}{2} + \eta_s\right)h_s(p) + \overline{\text{Flow}}(h_s) - m_{\sigma_s}^2 \dot{A}_s
	\,.
\end{align}
To take into account multi-meson scatterings in the strange sector, we utilise a full effective potential $V(\rho_l, \rho_s)$, where $\rho_s = \sigma_s^2 / 2$. To simplify the calculation, we do not fully resolve the two-dimensional effective potential, but use a Finite Element (FE) discretisation of the light meson part, and a Taylor expansion in $\rho_s$ around the running {EoM} ${\rho_{0,k} = (\rho^{(0)}_{l,k}, \rho^{(0)}_{s,k})}$. This expansion reads thus
\begin{align}
	V(\rho_l, \rho_s) \approx V_l(\rho_l) + \sum_{n=1}^{N_s} \frac{\lambda^{(s)}_{n,k}(\rho^{(0)}_{l,k})}{n!} (\rho_s - \rho^{(0)}_{s,k})^n
	\,,
\end{align}
where we take the order of the potential as $N_s = 5$. The flow of $\rho_{s,k}^{(0)}$ is deduced from the {EoM} of the strange~sector
\begin{align}
	\partial_{\sigma_s} V = \frac{1}{\sqrt{2}}c_{\sigma_s}
	\,,
\end{align}
as
\begin{equation}
	\partial_t \rho_{s,k}^{(0)} = \left\{
	-\frac{2 \rho_s\partial_t\partial_{\rho_s}V }{\partial_{\rho_s}V+2 \rho_s \partial_{\rho_s}^2V} - \eta_{\sigma_s}\rho_s
	\right\}_{\rho = \rho_{0,k}}
	\,.
\end{equation}
The flow of the coefficients $\lambda^{(s)}_{n,k}$ can be directly inferred from the flow of the effective potential,
\begin{align}
	\partial_t \lambda^{(s)}_{n,k} = \bigg\{&
	\partial_t\partial_{\rho_s}^n V
	+ \partial_{\rho_s}^n (\eta_{\sigma_s} \rho_s \partial_{\rho_s}V )
	\notag\\[1ex]
	&\qquad+ \lambda^{(s)}_{n+1,k}\partial_t \rho_s
	\bigg\}_{\rho = \rho_{0,k}}
	\,.
\end{align}
As argued in \cite{Ihssen:2024miv}, we can assume that the Taylor expansion is a valid discretisation scheme as long as we are far away from the chiral limit. For the strange sector, this is fulfilled in physical {QCD} due to the large current quark mass of the strange quark $m_s^0\approx\qty{93}{\MeV}$, which keeps the {EoM} sufficiently far away from the chiral limit.
On the other hand, we cannot fully rule out non-local effects at high densities. However, these should be present at even larger $\mu_B$ than for the light sector, due to the larger explicit symmetry breaking for the strange quark. Therefore, as long as we do not go to very high $\mu_B$ at low temperatures, we do not expect this to be of relevance.
Although we choose here $N_s = 5$, we have tested that no relevant difference occurs for orders $N_s \geq 3$.

%
\begin{table*}[t]
	\hspace{-5pt}
	\begin{tabular}{>{\centering}m{0.12\linewidth} ||>{\centering}m{0.06\linewidth}||>{\centering}m{0.16\linewidth}||>{\centering}m{0.16\linewidth}||>{\centering}m{0.16\linewidth}|| >{\centering\arraybackslash}m{0.16\linewidth}}
		Mass & Value & Parameter for $s_8$ & $r_8^{b=0,c=2}$ & $r_8^{b=1,c=1.5}$ & mod, $s_8$\\[1ex]
		\hline&&&&&\\[-2ex]
		$m_{\pi}$ [MeV] &  138 & $c_{\sigma_l} = \qty{7.5}{\GeV^3}$&$c_{\sigma_l} = \qty{6.9}{\GeV^3}$&$c_{\sigma_l} = \qty{7.0}{\GeV^3}$&$c_{\sigma_l} = \qty{7.4}{\GeV^3}$\\[1ex]
		\hline&&&&&\\[-2ex]
		$m_l$ [MeV]      &  $350$ & $a= 0.0582$ & $a= 0.04015$ & $a= 0.0350$ & $a= 0.0665$ \\[1ex]
		\hline&&&&&\\[-2ex] 
		$m_s$ [MeV]      &  $485$ & $c_{\sigma_s} = \qty{215}{\GeV^3}$&$c_{\sigma_s} = \qty{240}{\GeV^3}$&$c_{\sigma_s} = \qty{235}{\GeV^3}$&$c_{\sigma_s} = \qty{210}{\GeV^3}$\\[1ex]
	\end{tabular}
	\vspace{7pt}
	\caption{Initial conditions for our $N_f=2+1$ {QCD} calculation. The parameters in $\Gamma_{\Lambda_\textrm{UV}}$ are chosen such as to adjust the corresponding masses. 
	The quark mass is adjusted using an enhancement $\alpha$ of the light and strange quark-gluon vertices, see \cite{Fu:2019hdw, Ihssen:2024miv}. The last column contains the parameters for the setting with a modified gluon propagator input, see \Cref{sec:QCDPD_initial_detail}
		\hspace*{\fill}	
	}
	\label{tab:QCDPD_initial} 		
\end{table*}

\subsection{Glue sector}
\label{sec:QCDPD_setup_glue}

We parametrise the gluon propagator as
\begin{align}\label{eq:QCDPD_glueParam}
	[\Gamma_{AA}(p)]^{ab}_{\mu\nu} &= \tilde{Z}_A (p^2 + m_A^2)\Pi^\perp_{\mu\nu}(p)\delta^{ab} 
	\notag\\[1ex]
	&= Z_A p^2 \Pi^\perp_{\mu\nu}(p)\delta^{ab}
	\,,
\end{align}
i.e. we explicitly split off the gluon mass $m^2_A$ in our parametrisation to improve the thermodynamic treatment of the glue sector. While the parametrisation with $\tilde{Z}_A$ in \labelcref{eq:QCDPD_glueParam} is fed back into the dynamics, we show the usual wave-function $Z_A$ in our results, which can be directly reconstructed from the combination of $\tilde{Z}_A$ and $m^2_A$. As we take the identification $k = p$ in the glue sector, this reads explicitly for the $k$-dependence thereof,
\begin{equation}
	Z_{A,k} = \frac{\tilde{Z}_{A,k} (k^2 + m_{A,k}^2)}{k^2}
	\,.
\end{equation}
Using the {RG}-invariant formulation of \cite{Ihssen:2024miv}, we absorb $\tilde{Z}_A$ into the gluon field. The corresponding anomalous dimension is defined through the difference flow to the gluon vacuum input data $\tilde{\eta}_{A,N_f=2}^{(\textrm{input})}$,
\begin{align}
	\tilde{\eta}_A(\mu_B,T) = &\tilde{\eta}_{A,N_f=2}^{(\textrm{input})} - \tilde{\eta}_{A,N_f=2}^{(q)}(0,0) 
    \notag\\[1ex]
    &+ \tilde{\eta}_{A,N_f=2+1}^{(q)}(\mu_B,T)
	\,,
\end{align}
where the two last terms include respectively the $N_f=2$ quark loops and the $N_f=2+1$ quark loops. This approximation without back-coupling has been discussed before in \cite{Fu:2019hdw} and shown to yield good results.

\begin{figure}[b]
	\centering
		\hspace{-20pt}
		\includegraphics[width=1.06\linewidth]{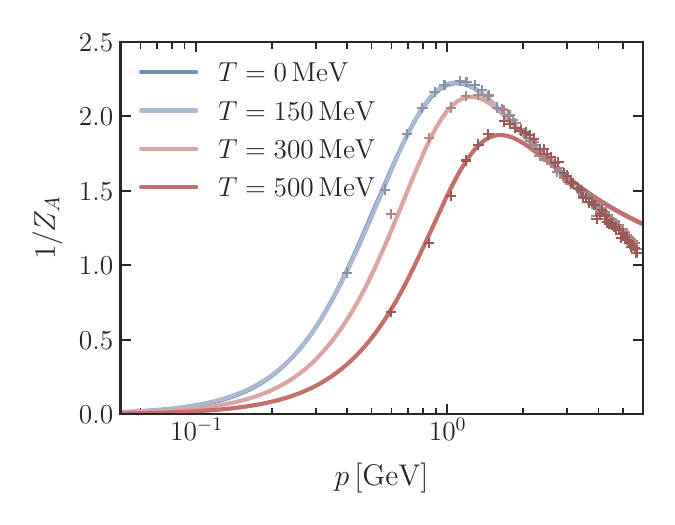}
		\caption{The glue propagator depending on temperature at $\mu_B=0$. We also show, for comparison, discrete lattice data points from an $N_f=2+1+1$ {QCD} calculation \cite{Ilgenfritz:2017kkp} at the depicted temperatures. Note that the gluon propagator sees almost no change between $T=0$ and $T=\qty{150}{\MeV}$.}
		\label{fig:gluonProps}
\end{figure}
The full calculation of the thermal gluon mass in our setup is impeded by the fact that our choice of infrared closure of the gauge fixing is affected by a finite temperature. The consequence thereof is that the difference flow to the $N_f=2$ vacuum result cannot be used without fully re-tuning the system to a scaling solution at every temperature.
Therefore, we do not take corrections between $N_f=2$ and $N_f=2+1$ for the gluon mass into account, and only add the thermal fluctuations. In other words, our glue mass reads
\begin{equation}
	m^2_A(\mu_B,T) = (m^2_{A,N_f=2})^{(\textrm{input})} + \Delta m^2_{A}(\mu_B,T)
	\,,
\end{equation}
and the flow of $\Delta m^2_{A}(\mu_B,T)$ is computed as
\begin{align}
	\partial_t\Delta m^2_{A}(\mu_B,T) =\, &\overline{\text{Flow}}^{(q)}(m^2_A)\Big\vert_{(\mu_B,T)} 
    \notag\\[1ex]
	&- \overline{\text{Flow}}^{(q)}(m^2_A)\Big\vert_{(0,0)} \notag\\[1ex]
	&+ \tilde{\eta}_A \Delta m^2_{A}(\mu,T)
	\,.
\end{align}
where the two flows both include all $N_f=2+1$ quark loops.
We show the resulting gluon propagators in \Cref{fig:gluonProps}, together with lattice data from \cite{Ilgenfritz:2017kkp}.

A crucial contribution from the glue sector at finite temperature is the Polyakov loop. We choose an effective inclusion of the Polyakov loop potential by using the pure Yang-Mills potential from \cite{Lo:2013hla} for the glue part of the Polyakov loop potential $V(L,\bar{L})$.
To this end, it is useful to parametrise the Polyakov loop using the dimensionless variable
\begin{align}
	\hat{\varphi} &= \frac{\beta\,g}{2\pi}A_0 = \hat\varphi_3 \lambda^3 + \hat\varphi_8 \lambda^8 \in \textrm{CSA}[\mathfrak{su}(N_c)]
	\,,
    \notag\\[1ex]
    L(A_0, \boldsymbol{x}) &= \frac{1}{N_c} \textrm{tr}\,\mathcal{P}e^{2\pi\imag\,\hat{\varphi}}
	\,.
\end{align}
We already made use of gauge-invariance to rotate $\hat{\varphi}$ into the Cartan subalgebra (CSA) of $\mathfrak{su}(N_c)$, spanned very explicitly by the two Gell-Man matrices $\lambda^3$ and $\lambda^8$. In particular, the set of shifts
\begin{equation*}
	\hat{\varphi} \xrightarrow{\mathfrak{z} \in \mathcal{Z}(\text{SU(3)})} \hat{\varphi} + \theta_\mathfrak{z}
	\,,
\end{equation*}
where $\mathcal{Z}(\text{SU(3)})$ is centre of SU(3), reads in this representation
\begin{equation}
	\theta_\mathfrak{z} \in \left\{
	0,\,\frac{2}{\sqrt{3}}\lambda^8,\,\lambda^3 - \frac{1}{\sqrt{3}} \lambda_8
	\right\}
	\,.
\end{equation}
\begin{figure}[t]
	\centering	
	\hspace{-5pt}\includegraphics[width=0.98\linewidth]{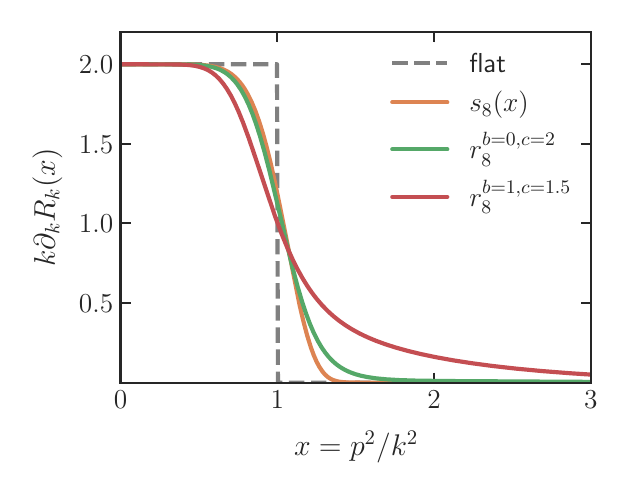}\vspace{-0.5em}
	\caption{We show the scale derivatives of the three regulator shape functions used for the determination of the systematic errors for the {QCD} temperature crossover. The phase diagram has been resolved using the shape function $s_8(x)$. For comparison, we also show the flat regulator shape. }
	\label{fig:QCDPD_Regulators}
\end{figure}
It follows that the Polyakov loop reads in terms of $\hat\varphi$
\begin{equation}\label{eq:QCD_PolyakovParam}
	L(\hat\varphi_3,\hat\varphi_8) = \frac13 \left(
	e^{-\frac{2\pi\imag\hat\varphi_8}{\sqrt{3}}} +
	2\,\cos(\pi\hat\varphi_3)e^{\frac{\pi\imag\hat\varphi_8}{\sqrt{3}}}
	\right)
	\,.
\end{equation}
At least at $\mu_B=0$, the extrema of the Polyakov-loop potential $V_\textrm{Pol}(L,\bar{L}) = V_\textrm{Pol}(\hat\varphi_3,\hat\varphi_8)$ are located at $\hat\varphi_8 = 0$. This is a consequence of charge conjugation symmetry and stops being valid at finite $\mu_B$. However, in this work we will use $\varphi_8 \approx 0$ as an approximation and suppose $L = \bar{L}$.
Note that \labelcref{eq:QCD_PolyakovParam} is helpful to translate between $L(A_0)$ and $A_0$.

This different parametrisation is chosen due to the value of $\langle L(A_0) \rangle$ being not easily accessible in functional approaches, as it represents a fully resummed, global quantity. In particular, accessing the associated Polyakov-loop potential $V_\textrm{Pol}(\langle L(A_0) \rangle)$ is not straightforward, but may be possible by field redefinition techniques, see \cite{Ihssen:2024ihp}. We note however, that as $L = L(A_0)$, the resolution of the potential directly in terms of $A_0$ is equivalent, 
\begin{equation}
	V_\textrm{Pol}(\langle L(A_0) \rangle) = V_\textrm{glue}(\langle A_0 \rangle)
	\,,
\end{equation}
which is the strategy we will follow here.

Together with this parametrisation, the pure Yang-Mills Polyakov loop potential $V_\textrm{YM}(L,\bar{L}; t_\textrm{YM})$ can be related to the glue contribution in full {QCD} by a rescaling of the reduced temperature $t_\textrm{YM}$, as has been found in \cite{Herbst:2015ona, Haas:2013qwp}. Explicitly, we take a rescaling as in \cite{Herbst:2015ona, Fu:2019hdw} and use
\begin{equation}
	t_\textrm{YM} \to t_\textrm{glue} = \alpha\,\frac{T-T_{c,\textrm{glue}}}{T_{c,\textrm{glue}}}
	\,,
\end{equation}
with $\alpha=0.57$ and $T_{c,\textrm{glue}} = \qty{240}{\MeV}$.
In our setup, we calculate the quark contributions to $V(L,\bar{L})$ explicitly and add them to the pure glue background from \cite{Lo:2013hla}.
To do so, we utilise one further approximation for the input data, which is 
$\langle L(A_0) \rangle = L(\langle A_0 \rangle)$. We calculate $V(L,\bar{L})$ on a discrete lattice and find the minimum at every time-step using a spline approximation to $V(L,\bar{L})$.

For the low-energy dynamics, the resulting expectation value $L$ at the minimum of is then fed back into the propagators of the theory as well as the flow of the mesonic potentials.

\section{Initial conditions}
\label{sec:QCDPD_initial_detail}

Our calculation is performed using three different regulators, taken from two classes of regulators optimised for convexity restoration \cite{Ihssen:2024miv, Sattler:2025hcg}.
We parametrise the regulators as 
\begin{equation}
    R^{(s/r)}_n(k,x=p^2/k^2)=k^2\,\begin{cases}
        r_n(x)\,,
        \\[1ex]
        s_n(x)\,,
    \end{cases}
\end{equation}
and the corresponding shape functions $r_n$ and $s_n$ are given by
\begin{align}
		r_n(x) &= \exp\left({-f_n(x)}\right)\,, \quad\text{where}\, f_n(x) = \frac{P_n(x)}{Q_{n-1}(x)}\,,
\notag\\[1ex]
    	s_n(x) &= \exp\left({-\sum_{i=1}^{n} \frac{x^i}{i}}\right)\,,
\end{align}
where the polynomials $P_n(x)$, $Q_{n-1}(x)$ are uniquely determined by convexity constraints, for the explicit form see \cite{Ihssen:2024miv, Sattler:2025hcg}. For the calculation, we use $s_8$, $r_8^{b=0,c=2}$ and $r_8^{b=1,c=1.5}$, as well as a setup with $s_8$ and modified glue input. For the modified glue input, we modify the gluon propagator's peak height by $\sim$ 0.5\%, consistent with the error estimate in \cite{Cyrol:2017qkl}.
Across all simulations, we take uniformly the initial meson mass $m_{\phi_l}^2 = m_{\sigma_s}^2 = 10^7\,\textrm{GeV}^2$ and an initial Yukawa coupling $h_l = h_s = 1$. Furthermore, the initial value of all avatars of the strong coupling is taken to be
\begin{equation}
	\alpha_{A^3} = \alpha_{A^4} = \alpha_{s\bar{s}A} = \alpha_{l\bar{l}A} = 0.216
	\,.
\end{equation}
We summarise all other, regulator-dependent initial parameters in \Cref{tab:QCDPD_initial}.

The explicit flows have been derived using \texttt{QMeS}\cite{Pawlowski:2021tkk} and the resulting diagrams traced with \texttt{FormTracer}\cite{Cyrol:2016zqb}. Tensorial bases and the projection operators for all vertices have been derived using \texttt{TensorBases}\cite{Braun:2025gvq} and the resulting flow equations solved in the high-performance C++ framework of \texttt{DiFfRG}\cite{Sattler:2024ozv}.

\bibliography{ref-lib}

\end{document}